\documentclass[11pt,preprint]{aastex}

\shorttitle{Globular Cluster Ages}
\shortauthors{De Angeli et al.}

\begin{document}

\title{GALACTIC GLOBULAR CLUSTER RELATIVE AGES. II.  
\footnote{Based on observations with the NASA/ESA {\it Hubble Space Telescope},   
obtained at the Space Telescope Science Institute, which is operated 
by AURA, Inc., under NASA contract NAS 5-26555, and on observations 
made at the European Southern Observatory, La Silla, Chile, and with 
the Isaac Newton Group Telescopes.}.} 

\author{Francesca De Angeli \altaffilmark{1}, Giampaolo Piotto \altaffilmark{1},
Santi Cassisi \altaffilmark{2}, Giorgia Busso \altaffilmark{3}}  
\author{Alejandra Recio-Blanco \altaffilmark{4}, Maurizio Salaris \altaffilmark{5}
Antonio Aparicio \altaffilmark{6}, Alfred Rosenberg \altaffilmark{6}} 

\altaffiltext{1}{Dipartimento di Astronomia, Universit\`a di Padova,  
vicolo dell'Osservatorio 2, I--35122 Padova, Italy; deangeli, piotto@pd.astro.it}
\altaffiltext{2}{INAF - Osservatorio Astronomico di Collurania, via Mentore Maggini, I--64100 Teramo, Italy;  
cassisi@te.astro.it}
\altaffiltext{3}{Institut f\"ur Theoretische Physik und Astrophysik, Universit\"at Kiel,  
Leibnizstrasse 15, D-24098 Kiel, Germany; busso@astrophysik.uni-kiel.de}
\altaffiltext{4}{Dpt. Cassiop\'ee, UMR 6202, Observatoire de la C\^ote d'Azur,
BP 4229, 06304 Nice Cedex 4, France; arecio@obs-nice.fr}
\altaffiltext{5}{Astrophysics Research Institute, Liverpool John Moores University,  
Twelve Quays House, Egerton Wharf, Birkenhead, CH41 1LD, UK;  ms@staru1.livjm.ac.uk}
\altaffiltext{6}{Instituto de Astrof\'{\i}sica de Canarias, Via Lactea, E-38200 La Laguna, Tenerife, 
Spain; antapaj, alf@iac.es}

\begin{abstract}
We present accurate relative ages for a sample of $55$ Galactic
globular clusters. The ages have been obtained by measuring the
difference between the horizontal branch and the turnoff in two,
internally photometrically homogeneous databases. 
The mutual consistency of the two data sets has been
assessed by comparing the ages of $16$ globular clusters in common
between the two databases. We have also investigated the consistency
of our relative age determination within the recent stellar model
framework.

All clusters with ${\rm [Fe/H]}<-1.7$ are found to be old, and coeval,
with the possible exception of two objects, which are marginally
younger. The age dispersion for the metal poor clusters is $0.6$ Gyr
(rms), consistent with a null age dispersion.  Intermediate
metallicity clusters ($-1.7<{\rm [Fe/H]}<-0.8$) are on average $1.5$ Gyr
younger than the metal poor ones, with an age dispersion of $1.0$ Gyr
(rms), and a total age range of $\sim 3$ Gyr. About $15\%$ of the
intermediate metallicity clusters are coeval with the oldest clusters. 
All the clusters with ${\rm [Fe/H]}>-0.8$ are $\sim 1$ Gyr
younger than the most metal poor ones, with a relatively small age
dispersion, though the metal rich sample is still too small to allow
firmer conclusions.  There is no correlation of the cluster age with
the Galactocentric distance. We briefly discuss the implication of these 
observational results for the formation history of the Galaxy.
\end{abstract}


\section{Introduction}

Galactic globular clusters (GCs) play a key role in stellar 
astrophysics and cosmology. Stars in a given GC are to a good 
approximation coeval, were born with the same initial chemical 
composition, and are located approximately at the same distance from 
the observer. These characteristics make them valuable benchmarks for 
testing stellar evolution theories through comparisons of theoretical 
isochrones and luminosity functions with their observational 
counterparts (see, e.g. \citealp{renzinifusipecci88}). In addition, 
due to the uniform age and initial chemical composition of stars belonging to 
a given cluster, the estimation of the cluster age is 
relatively straightforward, as shown in the pioneering work of   
\citet{sandageschwarzschild52}, \citet{hoyleschwarschild55}.  
 
Given that GCs are the oldest systems populating our Galaxy (and 
external galaxies) for which an age estimate is feasible, their ages 
provide a lower limit to the age of the Universe, whilst their 
relative ages (and possible correlations with metallicity, abundance 
patterns, and position) disclose fundamental information about 
Galactic formation mechanisms and timescales. Recent work on the 
age distribution of sizeable samples of GCs 
\citep{b98,salarisweiss98,rspa99,vandenberg00,salarisweiss02} 
has led to a broadly consistent scenario in which metal poor clusters 
up to some intermediate metallicities are largely coeval, whereas at higher 
${\rm [Fe/H]}$ values an age spread exists, with possibly an age-metallicity relation.  
The age distribution with respect to the position within the Galaxy 
shows that 
an age spread is present at all distances, without any correlation with the 
distance from the Galactic centre \citep{rspa99}. 
 
Unfortunately, absolute age measurements are still affected by a 
number of uncertainties, in particular by the large remaining errors on 
the GC distances and reddening \citep{grat03}.  Relative ages can be 
obtained with much higher accuracy by measuring the 
position of the main sequence turn-off (TO) -- which is the most-used 
age indicator -- relative to some other feature in the colour magnitude 
diagram (CMD) whose location is little or not at all dependent on age 
(Stetson, VandenBerg, \& Bolte 1996, Sarajedini, Chaboyer, \& Demarque 1997). 
However, any observed age indicator must be 
compared with theoretical predictions in order to derive an age. 
Uncertainties in both the 
input physics \citep{cha98}, and the transformation from the 
theoretical to the observational plane \citep{b98}, constitute a 
significant source of errors in the final age estimate, particularly 
when we want to compare ages of clusters with different metal 
content. In the present work, in order to minimise the uncertainties coming 
from the models, we used an updated set of theoretical tracks. More specifically, the 
theoretical isochrones employed in our analysis are the 
$\alpha$-enhanced counterparts 
of the \citet{pietrinferni04} model and isochrone library, and have
already been discussed briefly in \citet{cassisi04}. The reliability and accuracy 
of this theoretical framework has already been tested by comparison with various  
empirical data sets by \citet{riello03} and by \citet{salaris04}. 
 
In the measurement of accurate relative ages, the photometric 
homogeneity of the data, and the homogeneity in the methods used to 
measure the age indicator parameters, are of fundamental importance, as 
we showed in \citet{rspa99}, where, for the first time, we have been 
able to measure relative ages on a truly photometrically homogeneous 
database for a large sample (34) of GCs. In this paper we intend to 
extend the work of \citet{rspa99}, by adding a new, larger,  
photometrically 
homogeneous catalogue of CMDs, i.e. the HST snapshot catalogue by 
\citet{snapshot}. For methodological consistency, we have also 
re-measured all the parameters relevant for an age estimate in the 
CMDs of \citet{r00a,r00b} used by \citet{rspa99}. 
 
The age-dating method used in our analysis is based on the magnitude 
difference $\Delta V_{TO-HB}$ (in the $F555W$ band for the HST data, and 
in the analogous $V$-Johnson band for the goundbased ones) between the TO and 
the horizontal branch (HB), i.e. the so called vertical method 
\citep{svb96, scd97}.  The TO brightness is the age indicator 
(the TO gets fainter for increasing age) whereas the HB level is 
unaffected in the age range typical of GCs; this implies that older 
clusters display a larger value of $\Delta V_{TO-HB}$. For ages around 
$10$~Gyr the parameter $\Delta V_{TO-HB}$ scales approximately as 
$\delta \Delta V_{TO-HB}/\delta t \sim 0.1 \ {\rm mag} \ {\rm 
Gyr^{-1}}$. 
Since it is a differential quantity, the comparison between observed and 
predicted $\Delta V_{TO-HB}$ values is unaffected by uncertainties in 
cluster reddening and distance modulus estimates, superadiabatic convection treatment in 
the stellar models as well as uncertainties in the colour-effective temperature 
relation adopted for transferring the models from the theoretical plane to the observational one. 
Another advantage of using the $\Delta V_{TO-HB}$ parameter 
as age indicator is that it is weakly sensitive to uncertainties in 
the cluster ${\rm [Fe/H]}$, because at a fixed age the TO magnitude scales 
with ${\rm [Fe/H]}$ almost like the HB level. 

The plan of this paper is as follows. Section~2 describes the 
observational data, the measurement of the $\Delta V_{TO-HB}$ from the 
observed CMDs, and the estimate of the associated errors.  
Section~3 is devoted to the estimate of the age distribution of our 
cluster sample. A discussion of the main results is in Section~4. 
Conclusions are in Section~5.

\section{Measurement of the observational parameters}

\subsection{The databases} 

The present investigation is based on two databases of GC CMDs: 
\begin{itemize} 
\item the HST snapshot database of \citet{snapshot}; 
\item the groundbased database of \citet{r00a,r00b}. 
\end{itemize} 

Each of the two databases is internally photometrically homogeneous. 
All the CMDs from both catalogues can be found at the Padova Globular 
Cluster Group web page \\
\noindent (\url{http://dipastro.pd.astro.it/globulars}). 
 
Here, for the first time, we present an age estimate based on the HST 
snapshot catalogue.  This is a catalogue of CMDs for $74$ Galactic GCs, 
observed in their central regions with the WFPC2/HST camera in the 
photometric bands $F439W$ and $F555W$ (similar to the $B$, $V$ in the 
Johnson-Cousins photometric system).  For details about the 
observations, the reduction, and the photometric measurements please 
refer to \citet{snapshot}.   
  
The groundbased database has already been used for relative age 
measurements by \citet{rspa99}, who selected the best CMDs from the 
two, photometrically homogeneous data sets collected, respectively, at 
the ESO/Dutch telescope (for the southern sky GCs) by \citet{r00b}, 
and at the Isaac Newton Group Jacobus Kapteyn telescope (for the 
northern GCs) by \citet{r00a}. For details about the 
observations, the reduction, and the photometric measurements please 
refer to \citet{r00a,r00b}.   
 
Forty one out of $74$ clusters from the HST snapshot catalogue, and $30$ out 
of $52$ clusters from the groundbased catalogues, have a CMD which can be 
used for an age measurement. 
The remaining objects were excluded for 
several reasons: too shallow photometry, large field 
contamination, small number of member stars.
Sixteen clusters are in common between the 
two catalogues, which allowed us to check the consistency of the ages 
derived from the two databases. Combining the two 
catalogues, we have been able to estimate ages for $55$ GCs, 
representing more than $35$\%
of the Galactic GC population. These GCs cover 
a metallicity interval $-2.3<{\rm [Fe/H]}<-0.3$, and are located from the very 
central part of the Galaxy out to $\sim30$ kpc, therefore extending the 
metallicity and Galactocentric distance coverage of the original \citet{rspa99} 
paper. 

In our attempt to be as homogeneous as possible, and consistent with 
our previous investigations based on the HST snapshot catalogue, we have 
adopted the metallicities listed in \citet{rutledge97} calibrated over 
both the \citet[hereafter CG]{cg97}, as extended by \citet{carretta01},  
and \citet[hereafter ZW]{zw84} metallicity scales.  
In this paper, we will present the results obtained adopting both 
metallicity scales. For those clusters that do not appear in the 
\citet{rutledge97} catalogue, we adopted the original ZW value when available. 
Otherwise, we calculated the ZW metallicity from \citet[in the 
revised version of 2003]{harris} by fitting a straight line to the 
relation between the two scales. 
For clusters not in the \citet{rutledge97} compilation, ${\rm [Fe/H]}$ values
in the CG system were calculated from ZW values using the following equation 
\citep{carretta01}:
${\rm [Fe/H]}_{CG} = 0.61 + 3.04 {\rm [Fe/H]}_{ZW} + 1.981 {\rm [Fe/H]}_{ZW}^2 + 0.532 {\rm [Fe/H]}_{ZW}^3$. 
 
All the other cluster parameters used in this paper have been extracted 
from \citet[in the revised version of 2003]{harris}. 

\subsection{Measurement procedures}  
 
The relevant age indicator parameters described below have been 
extracted from both databases. It is worth emphasising the fact that the HST 
snapshot CMDs are in the $F439W$, $F555W$ WFPC2 flight system, while the 
groundbased CMDs are in the $V$, $I$ Johnson-Cousins photometric 
system. Therefore, the vertical parameters obtained from the two 
databases cannot be directly compared.  First, they must be compared 
with the theoretical models, transformed into the appropriate 
observational plane, for the age determinations. Only these final ages 
can be compared. 
  
In order to apply the vertical method, we need to measure the 
magnitude of the TO point and the Zero Age Horizontal Branch 
(ZAHB) level.   
 
For the TO measurement, we proceeded in the following way. First, 
the CMD of each clusters was cleaned by selecting stars based 
on the DAOPHOT PSF fitting parameters (CHI, SHA) and
their photometric errors. Second,
we extracted the fiducial main sequence (MS) for each cluster by taking 
the median of the colour distributions obtained in magnitude boxes 
containing a fixed number of stars, ranging from $40$ to $250$, depending 
on the total number of MS stars. A preliminary TO was then defined as 
the bluest point of the MS fiducial lines. For each cluster, we 
extracted a number of different MS fiducial lines by using different 
numbers of stars per box, and for each fiducial MS we associated a 
preliminary TO to its bluest point. Finally, we adopted as TO 
magnitude the mean value of all the preliminary TOs.  The error on the 
mean of the preliminary TO magnitudes gave a first estimate of the 
error associated to the adopted TO magnitude. The TO magnitude error 
was then better quantified following the procedure explained in the 
next Section. 
 
For the ZAHB magnitudes of the HST snapshot clusters, we adopted the 
values recently published by \citet{recioblanco05}. They present 
distance moduli and reddening estimates for $72$ Galactic globular 
clusters, based on the same photometric catalogue we are using here. The 
ZAHB magnitudes were estimated starting from the RR~Lyrae level, for 
low and intermediate metallicity clusters, and from the fainter 
envelope of the red HB, for metal rich ones. For a detailed 
description of the adopted method, please refer to 
\citet{recioblanco05}.  Here it suffices to say briefly that: 
\begin{itemize}  
\item  the ZAHB level for {\it metal rich clusters} (${\rm [Fe/H]}\geq-1.0$)  
was set as the magnitude of the lower envelope of the HB minus $3$ 
times the photometric error (resulting from artificial star 
experiments) at that magnitude; 
\item for {\it low and intermediate metallicity clusters} (${\rm [Fe/H]}<-1.0$) , 
a template cluster with similar metallicity and with a sizeable population 
of RR~Lyrae stars was selected from the literature. Five different 
templates were used to cover the entire metallicity range.  The CMD of the 
template cluster was shifted in colour and magnitude until its HB 
overlapped the HB of the cluster whose ZAHB level was to be 
determined. This procedure allowed to obtain the mean apparent 
magnitude of the cluster RR~Lyrae stars which was then transformed into 
the apparent ZAHB magnitude by using the relation \citep{c97}: 
\begin{equation}  
V_{\rm ZAHB}=V_{\rm RR~Lyrae} + 0.152 + 0.041 {\rm [M/H]}.  
\end{equation} 
\end{itemize}  
 
For methodological consistency, we repeated the measures of the TO and 
ZAHB magnitudes also for the groundbased CMDs. The TO measurement method by 
\citet{rspa99} is very similar to what we did in the present paper, 
while the approach to obtaining the ZAHB magnitudes is completely 
different.  The new ZAHB magnitudes for the groundbased CMDs have been 
measured exactly as described above for the HST data, and in more 
detail by \citet{recioblanco05}.  For the groundbased low and 
intermediate metallicity clusters we used the same templates as for the HST 
ones. However, the reference CMDs are different, as the groundbased 
catalogue is in the $V$, $I$ photometric system. Table~\ref{ground} 
lists the reference clusters, the metallicity range of the comparison 
clusters, and the source of the photometry for the reference 
CMDs.  
  
\begin{figure}
\includegraphics[width=1.0\textwidth,angle=0]{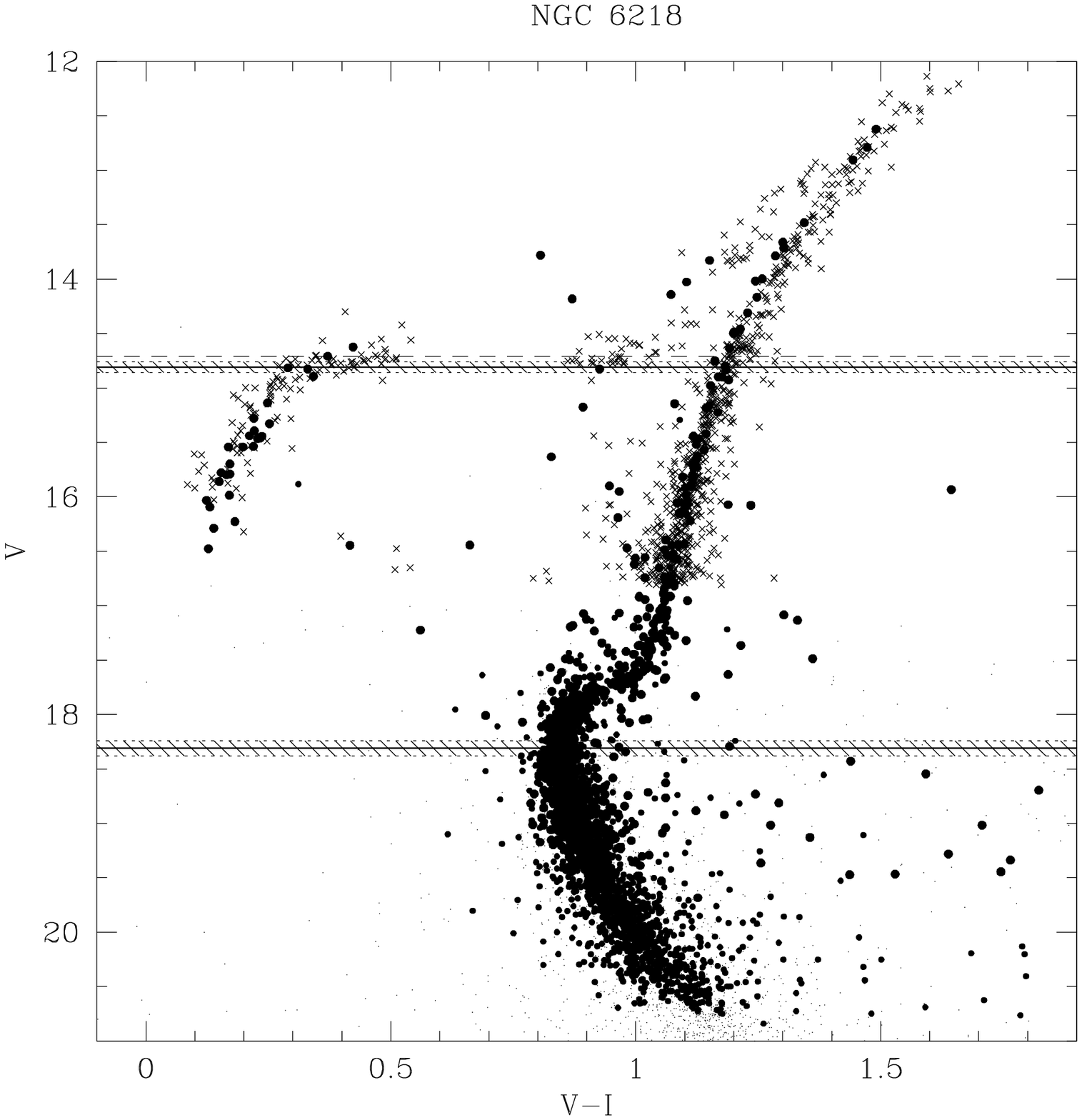}             
\caption{Groundbased CMD of the cluster NGC~6218 (dots). The crosses
are the brighter section of the CMD of NGC~5904, the template used
in the ZAHB measurements in the metallicity range of NGC~6218.
The horizontal solid lines are the TO and ZAHB magnitudes. The dashed
regions represent the 1-$\sigma$ uncertainties. The dashed line is the
RR~Lyrae mean magnitude of the template cluster. \label{fig0a}}
\end{figure}

\begin{figure}
\includegraphics[width=1.0\textwidth,angle=0]{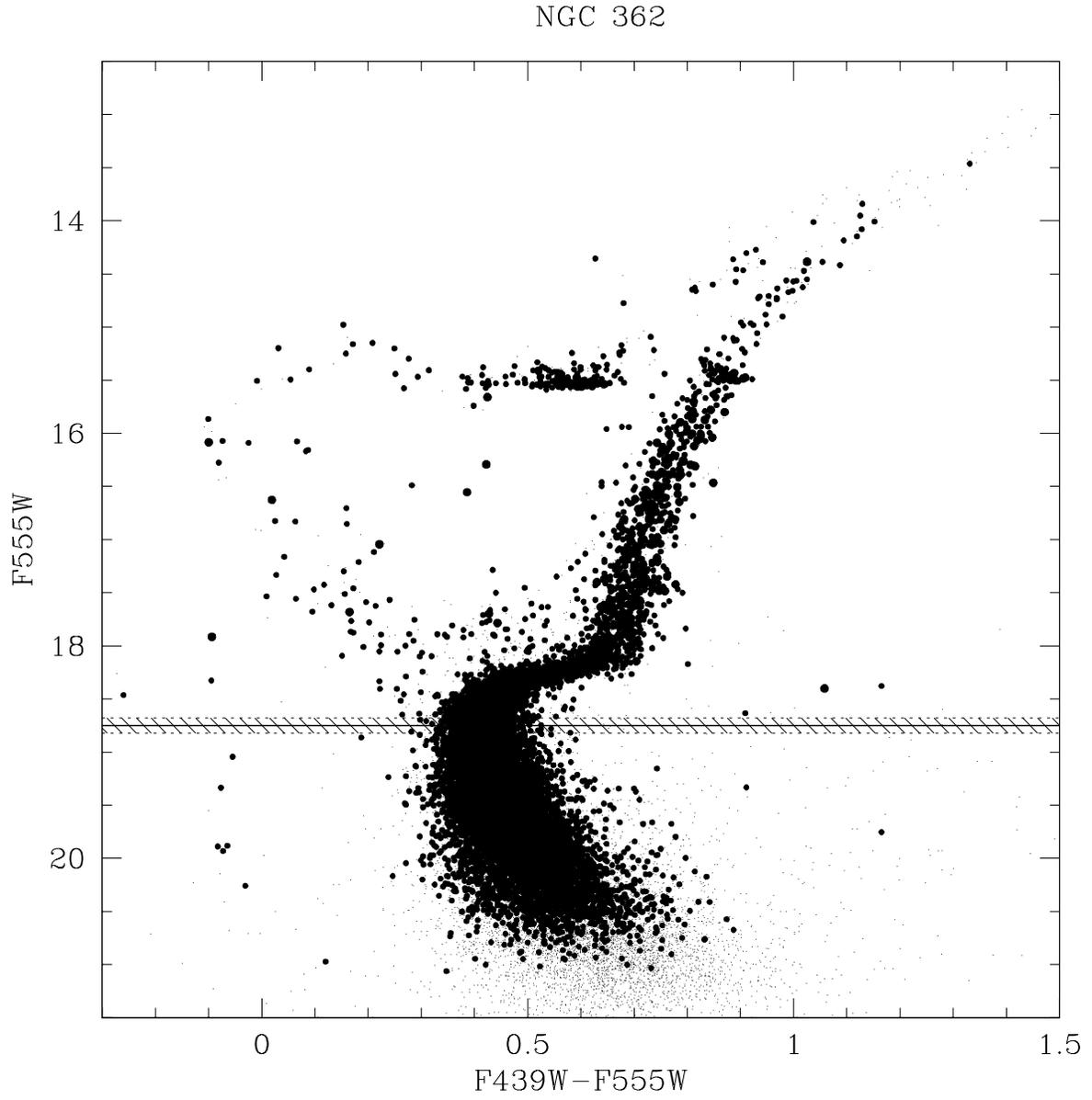}             
\caption{HST snapshot CMD of the cluster NGC~362. 
The horizontal solid line corresponds to the TO magnitudes. The dashed
region represents the 1-$\sigma$ uncertainty.\label{fig0b}}
\end{figure}

Figures~\ref{fig0a} and \ref{fig0b} show two examples of our 
results. The CMDs of two clusters are shown, one from the groundbased 
catalogue (NGC~6218) and one from the the snapshot database (NGC~362). 
In Fig.~\ref{fig0a}, the brighter part of the CMD of one of the templates
used for the ZAHB measurements is also shown. The template is NGC~5904.
NGC~6218 has a quite scarcely populated HB, nevertheless by applying our 
technique we were able to estimate the ZAHB magnitude. 
We would like to emphasise the fact that NGC~6218 is also contained in 
the snapshot catalogue and ZAHB measurements were made on the HST CMD and
published by \citet{recioblanco05}. Those measurements are in excellent 
agreement with those obtained from the groundbased CMD, considering the
different but similar bandpasses.

\subsection{Observational errors}  
  
In order to estimate the uncertainty in the adopted TO  
magnitude, we used the same method as in \citet{rspa99}. We built about 
one hundred synthetic CMDs for each cluster, using the isochrones by 
\citet{pietrinferni04}. The synthetic CMDs were constructed by 
adopting for each cluster the corresponding metallicity, an age in the range 
between $10$ and $13$ Gyr, the photometric errors (as estimated from 
artificial star experiments), and the total number of stars in the 
observed CMD, and varying only the initial random number generator 
seed. The same procedure used to determine the TO in the observed CMD 
(see the explanation in the previous Section) was then applied to the 
synthetic diagrams. 
We estimated the error in the TO colour and magnitude 
in each observed CMD as the standard deviation of all the TOs measured 
in the corresponding synthetic diagrams. 

This procedure does not take into account the effects due to differential 
reddening. Therefore the error estimated for clusters whose CMDs show
this effect (namely NGC~4372, NGC~5927, NGC~6273, and NGC~6544) is likely to 
be underestimated. The error on the mean of the preliminary TO magnitudes
for NGC~6273 (the worst case) is approximately a factor 2 larger than the 
error calculated from the synthetic diagrams. Nevertheless, we decided to 
adopt this technique for its reliability and for consistency with 
\citet{rspa99}.

Regarding the errors on the ZAHB magnitude of the intermediate and 
metal poor clusters, they are dominated by the error in matching the 
HB of the template cluster with that of the object cluster, 
particularly for clusters with only a blue HB.  As in 
\citet{recioblanco05}, we repeated the fit many times, and estimated 
the error as the maximum difference between the single shift 
values and the final adopted one.   
This error estimate has been 
divided by $3$, in order to have the classical 1-$\sigma$ value for a 
Gaussian distribution of the errors (in order to make it 
comparable with the errors of the TO magnitudes\footnote{Note that in 
Table~3 of \citet{recioblanco05} we list the maximum error.}). 
For the metal rich clusters, the main source of error is the uncertainty in 
the position of the lower envelope, particularly in clusters with a 
small number of stars. 
  
In estimating errors in the vertical parameter, we considered the 
errors on the ZAHB and TO magnitudes to be independent, and thus 
summed in quadrature the two contributions. 
 
\subsection{The measured parameters} 
  
Table~\ref{tab1} lists the most relevant cluster 
parameters, including those measured in this paper, for the HST 
snapshot clusters.  Col.~1 gives the cluster 
name; Col.~2 and 3 give the metallicity in the ZW and 
CG scale, respectively; Col.~4 gives the Galactocentric 
distance. Columns~5, 6 of Table~\ref{tab1} give the TO magnitudes and 
errors, respectively, in the $F555W$ band, while the $F555W$ ZAHB 
magnitudes and errors are in Col.~7, 8.  Finally, Col.~9, 10 
list the vertical parameter value, in the $F555W$ band, 
for the snapshot clusters.  
Table~\ref{tab2} provides the same quantities as Table~\ref{tab1}, 
but for the groundbased clusters. 
The magnitudes of Table~\ref{tab2} are therefore in the $V$-band. 
                              
\begin{figure}
\includegraphics[width=1.0\textwidth,angle=0]{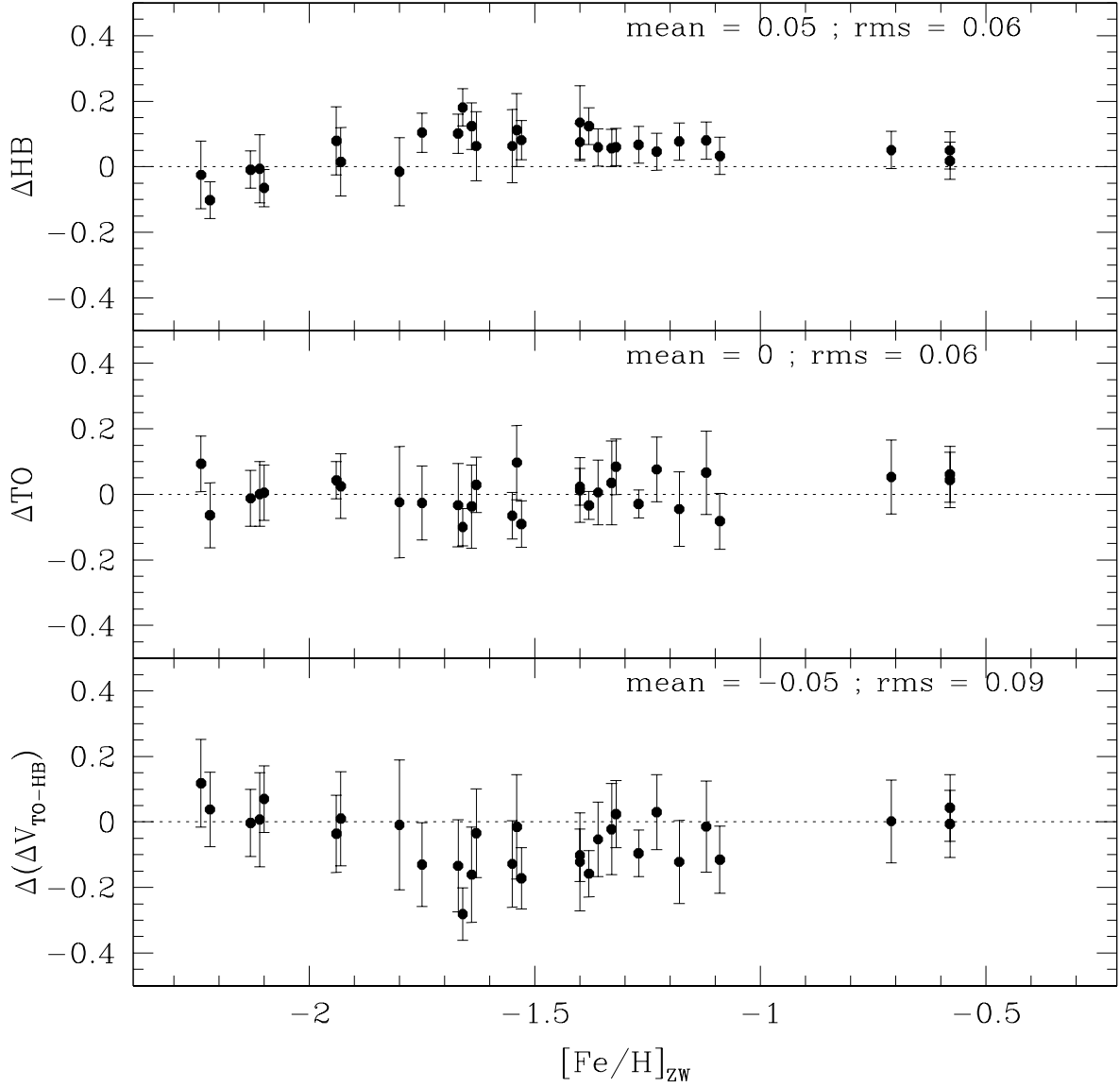}      
\caption{Comparison between our ZAHB ({\it upper 
panel}) and TO ({\it middle panel}) magnitudes and those listed by 
\citet{rspa99}. The {\it lower panel} shows the corresponding 
differences in the vertical parameters.\label{confr.gbn-gba}}
\end{figure}

As already mentioned, we have used the same database \citep{r00a,r00b} used by 
\citet{rspa99}, but for consistency reasons we decided to 
re-measure the vertical parameter. Figure~\ref{confr.gbn-gba} shows 
the comparison between the TO and ZAHB magnitudes, and between the 
vertical parameters of the present study and those listed by 
\citet{rspa99} for the groundbased clusters. While the TO values are in 
very good agreement, there is some hint of a 
trend with the metallicity (though it is mostly within the average 
measurement errors) for the differences in ZAHB magnitudes in the 
intermediate and low metallicity range. This was expected because of 
the different method adopted in the two studies and the different 
definition of the ZAHB level.  
\citet{rspa99} used the same (empirical) template for all the 
intermediate and metal poor clusters.  We think that the method used 
in the present paper is more reliable, as it is able to better account 
for the differences in the HB at varying the cluster metallicities. 
 
\section{Ages}  
  
\subsection{The age determination} 
 
To measure the cluster relative ages, we compared our observed 
vertical parameters with the theoretical counterpart 
obtained from the set of $\alpha$-enhanced isochrones of 
\citet{cassisi04}, available in both the WFPC2/HST photometric bands, 
and the Johnson-Cousins system. 
 
For each isochrone, the theoretical TO was determined as the bluest 
point of the MS, and the ZAHB level was taken as the magnitude of the 
ZAHB model at $\log {\rm T_{eff}}=3.85$.  We performed a 
$2$-dimensional bicubic spline to set a finer grid in age and 
metallicity, and thus to give an estimate of the age for each cluster 
in the sample. 
  
\begin{figure}
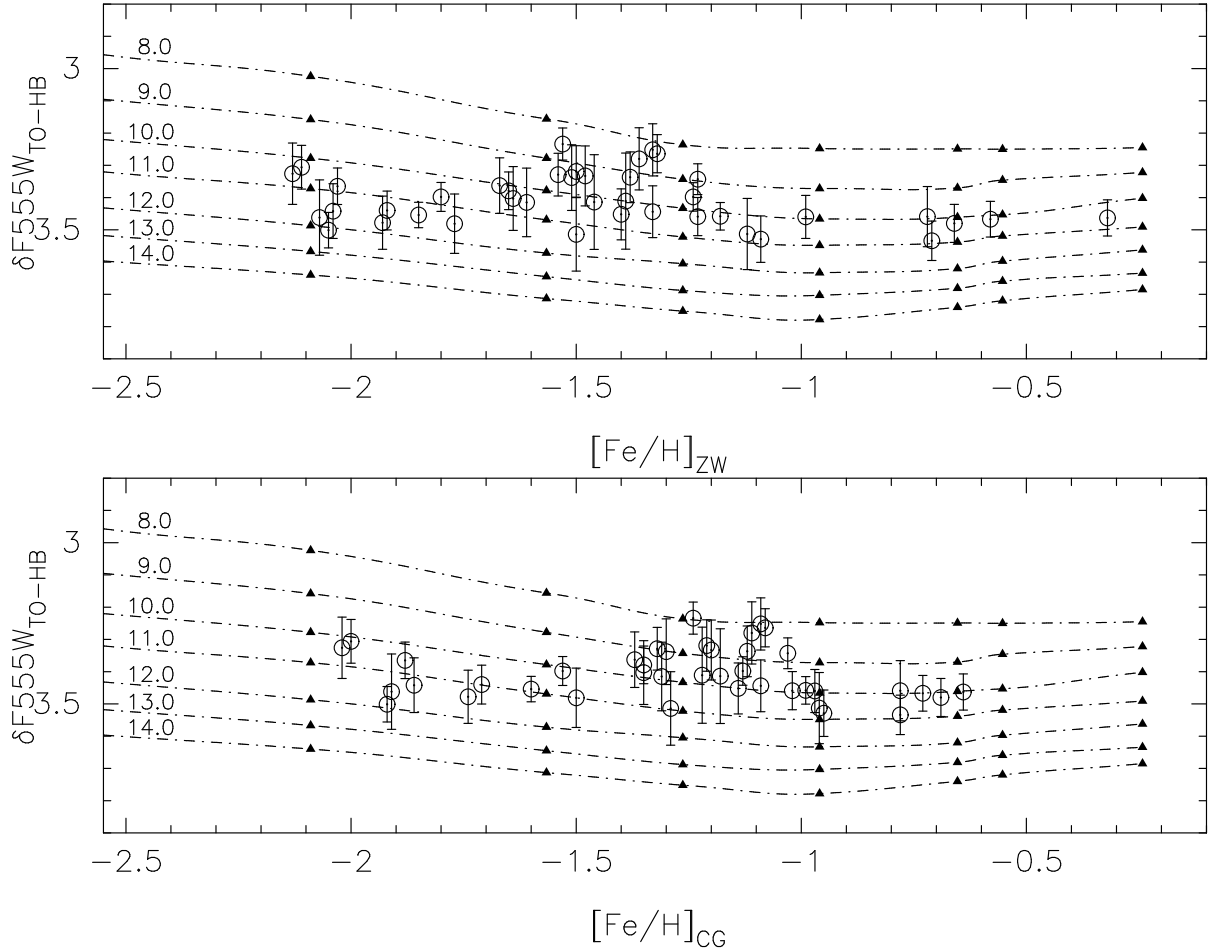

\includegraphics[width=0.38\textwidth,angle=270]{figure4a.eps}\\  
\includegraphics[width=0.38\textwidth,angle=270]{figure4b.eps}  
\caption{The vertical parameters measured on the HST snapshot cluster CMDs 
are plotted versus the metallicity (adopting the ZW scale 
in the {\it upper panel} and the CG one in the {\it 
lower panel}). The dashed lines show the theoretical predictions.  The 
isochrones are spaced by 1 Gyr (starting from 14.0 Gyr at the bottom).\label{c04aesnap}} 
\end{figure}

\begin{figure}
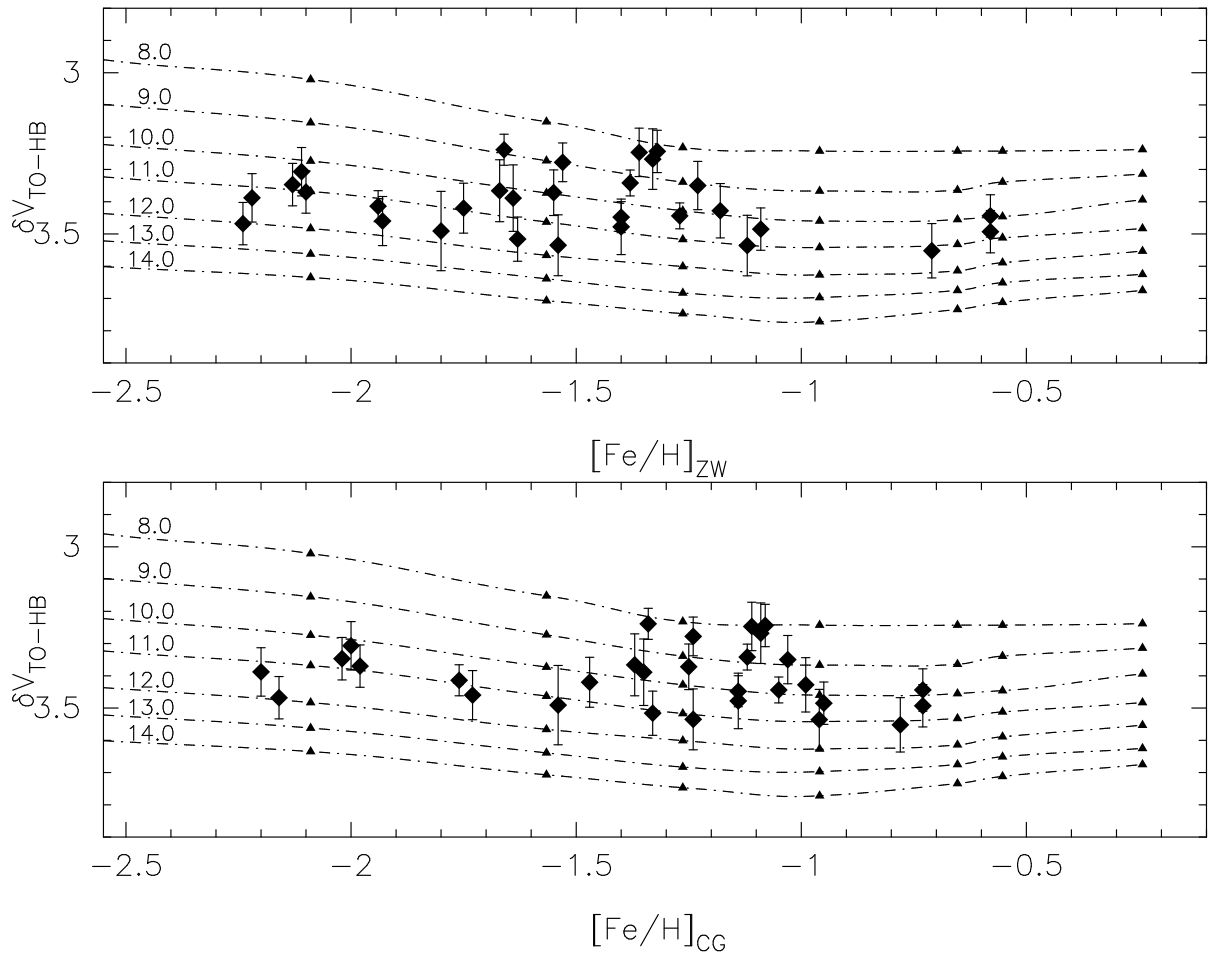

\includegraphics[width=0.38\textwidth,angle=270]{figure5a.eps}\\  
\includegraphics[width=0.38\textwidth,angle=270]{figure5b.eps}  
\caption{As in Fig.~\ref{c04aesnap} for the groundbased sample.\label{c04aegb}} 
\end{figure}

Figures \ref{c04aesnap} and \ref{c04aegb} show the theoretical $\Delta 
F555W_{TO-HB}$ and $\Delta V_{TO-HB}$ values as a function of ${\rm [Fe/H]}$ 
for various ages, together with the observational points from the 
HST snapshot and groundbased observations, respectively. The upper 
panel of each figure adopts  the ZW metallicity scale, 
whereas the lower panel refers to the CG 
scale.  The two data sets are presented separately due to the different 
photometric systems.   
 
Both Figures \ref{c04aesnap} and \ref{c04aegb} show that the mean 
age of the metal poor clusters is about $1.5$~Gyr older than the mean age 
of the clusters with ${\rm [Fe/H]}_{\rm ZW}>-1.7$ (${\rm [Fe/H]}_{\rm CG}>-1.4$). 
At intermediate metallicities ($-1.7<{\rm [Fe/H]}_{\rm ZW} <-0.8$), the age 
spread is increased by a group of a few clusters showing a younger 
age. This will be discussed further in the next Section. 
 
Keeping in mind that the purpose of this work is to measure relative 
ages, and following \citet{rspa99}, we calculated for each cluster a 
{\bf normalised age}, defined as the ratio between the cluster 
age and the mean age of the group of metal poor clusters. More 
specifically, in case of the ZW metallicity, we used as normalising 
factor the mean age of the clusters with ${\rm [Fe/H]}_{\rm ZW}<-1.7$, 
i.e. $11.2$ Gyr. For the CG metallicity scale, the normalising age is equal to 
$10.9$ Gyr, corresponding to the mean age of the clusters with 
${\rm [Fe/H]}_{\rm CG}<-1.4$.  Columns 11--14 of Tables~\ref{tab1} and \ref{tab2} give the 
estimated normalised ages and errors for the HST snapshot and the groundbased 
sample, respectively. In both tables, 
Col.~11, 12 give the normalised ages, and the corresponding errors adopting 
the ZW metallicity scale, and Col.~13, 14 the normalised ages 
and the corresponding errors for the CG metallicity scale. 
 
\begin{figure}
\includegraphics[width=1.0\textwidth,angle=0]{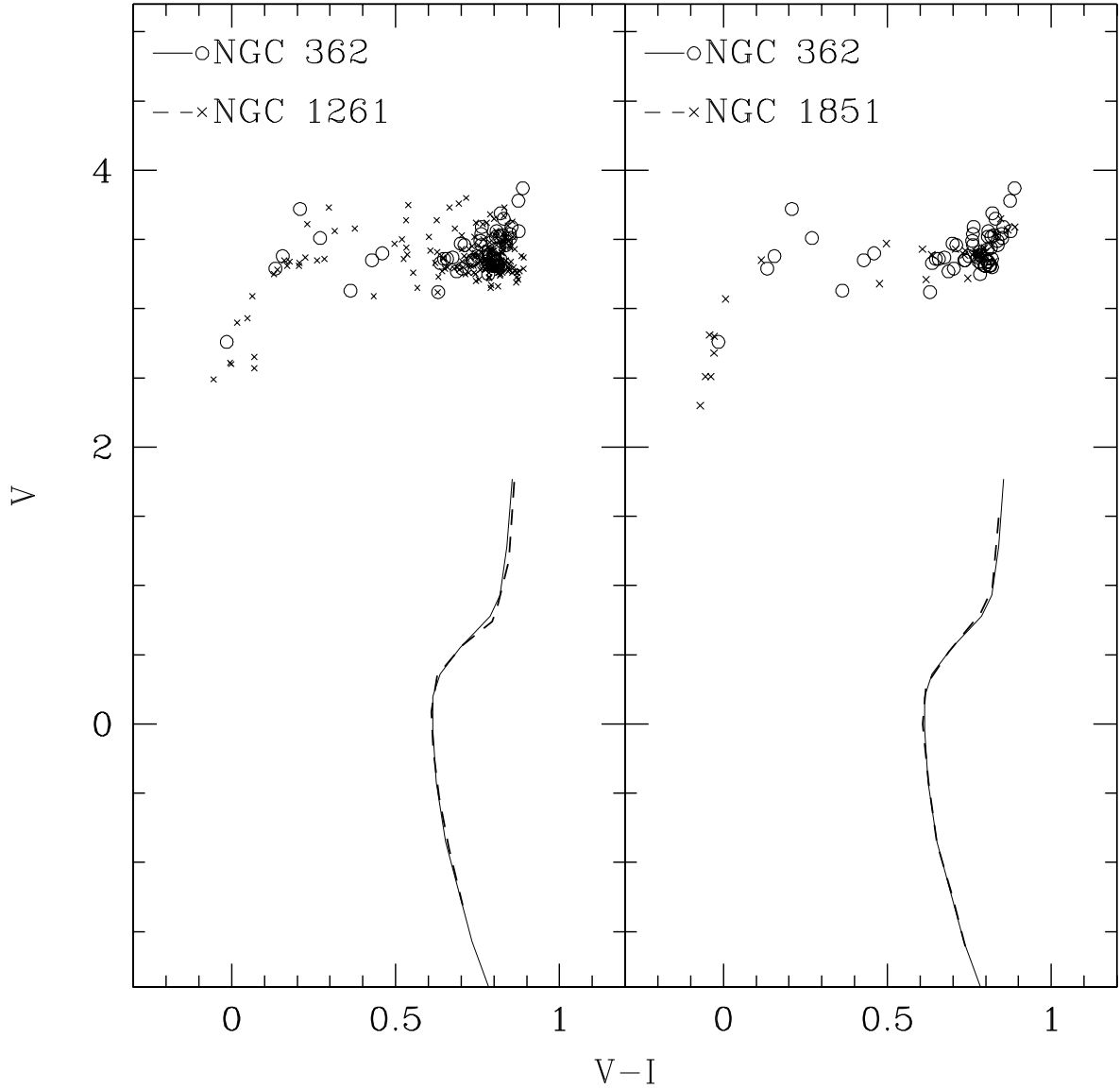}  
\caption{Comparison between the HB populations and the fiducial points reproducing 
the location of the main sequence and SGB of the CMDs of two couples 
of nearly coeval clusters: NGC~362 - NGC~1261 ({\it left panel}) and NGC~362 - NGC~1851
({\it right panel}). The HB stars and the fiducial points have been shifted in order 
to match the positions of the TOs.\label{comp_ml1}} 
\end{figure}

\begin{figure}
\includegraphics[width=1.0\textwidth,angle=0]{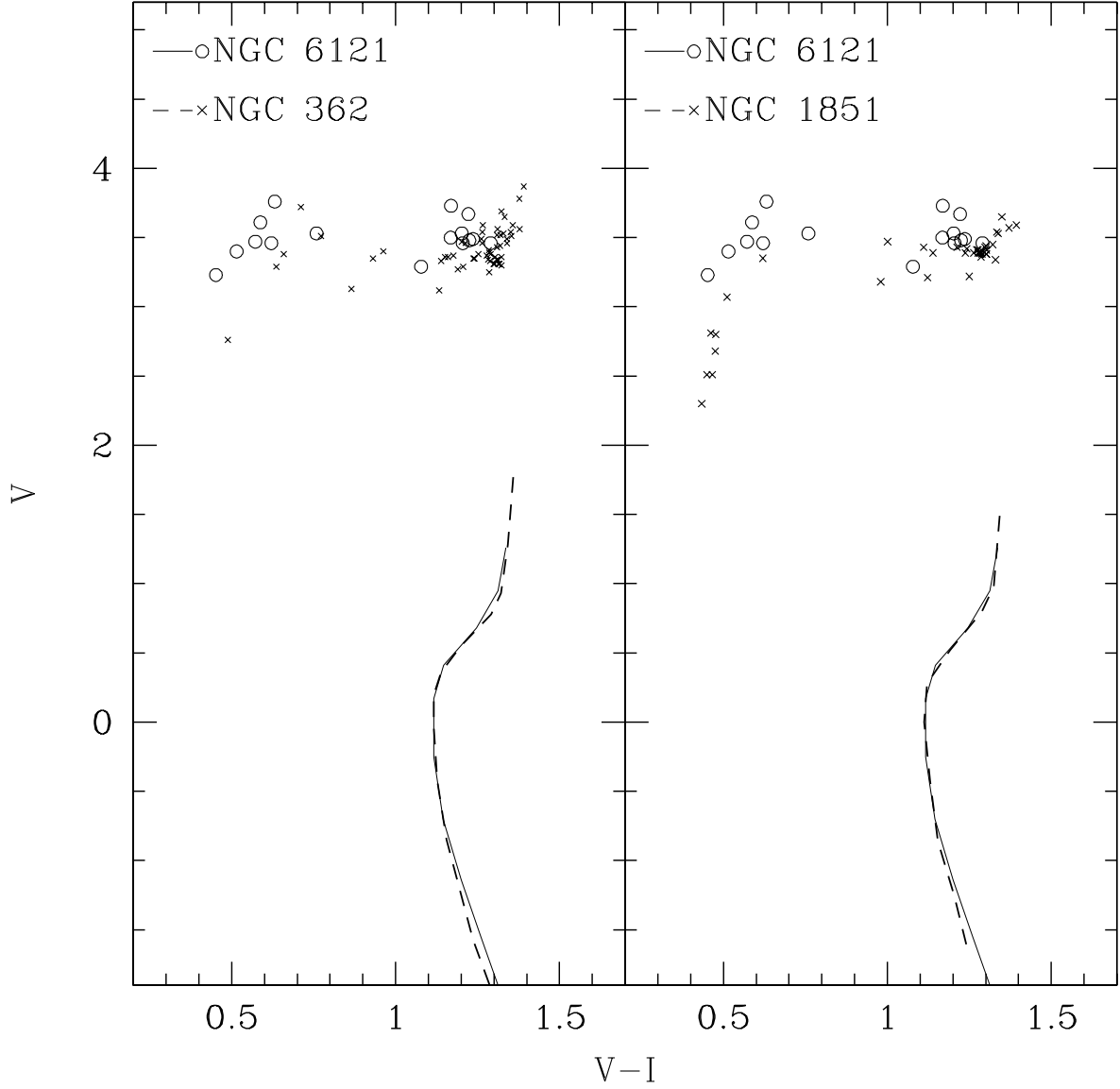}  
\caption{Comparison between the HB populations and the fiducial points reproducing 
the location of the main sequence and SGB of the CMDs of two young 
and nearly coeval clusters (NGC~362 and NGC~1851, {\it left} and {\it right panel} respectively) 
with the older NGC~6121. The HB stars and the fiducial points have been shifted in order 
to match the positions of the TOs.\label{comp_ml2}} 
\end{figure}

As it can be inferred from the values in Tables \ref{tab1} and \ref{tab2}, 
and as it will be discussed in Section 4, the group of clusters at
intermediate metallicity shows a large age dispersion.
We have verified that the age differences derived from the
vertical parameter are indeed correct.
In particular, we focussed our attention on a few clusters with
nearly the same metallicity, within $0.1$ dex, at ${\rm [Fe/H]}_{\rm ZW}\sim-1.3$.
Some of these clusters have been the subject of a long and lively
discussion in the literature in the last 10 years or so 
\citep[e.g.][]{svb96,sandquist96,vandenberg00} 
in the attempt to establish whether they are coeval or not.
According to the vertical parameter, NGC~362, NGC~1261, NGC~1851,
NGC~2808, NGC~5904 have all the same age, within the uncertainties
(5-8\%), and are 10-15\% younger than another group of older
(and coeval among them) clusters which includes NGC~288, NGC~5946,
NGC~6218, NGC~6121, NGC~6266, NGC~6362, NGC~6717.
In order to check this observational evidence, in
Figures~\ref{comp_ml1} and \ref{comp_ml2} we directly compare the
HB population and the fiducial points reproducing the location of the main sequence 
and sub-giant branch (SGB) of the CMDs of four couples of clusters selected among 
the above mentioned objects.
The HB stars and the fiducial points have been shifted in magnitude in order to have 
the TO point at magnitude 0. One of the two clusters has
been shifted in colour by the amount required to take the TOs at the same colour. 
Figure~\ref{comp_ml1} shows this comparison for two couples of nearly coeval clusters
(NGC~362 - NGC~1261 and NGC~362 - NGC~1851): all the sequences match
remarkably well, and it is hard to infer any age difference among these objects. 
These clusters were also found to be nearly coeval by \citet{sandquist96} and \citet{vandenberg00}.
Figure~\ref{comp_ml2} compares two of these young clusters (NGC~362 and NGC~1851) with NGC~6121,
which results older: despite the TO and SGB regions match well, the HB levels significantly
differ, as expected from our vertical parameters.
The difference in magnitude between the HB levels is indicative of the different ages.

\subsection{Comparison between the two catalogues} 
 
\begin{figure}
\includegraphics[width=1.0\textwidth,angle=0]{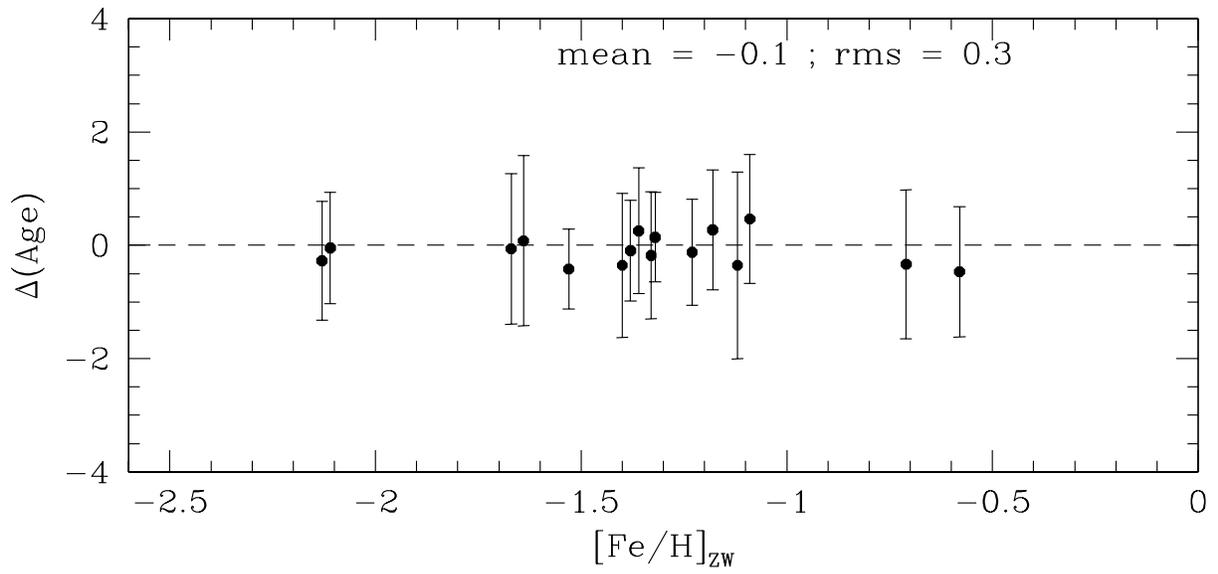}      
\caption{The age difference (in Gyr) from the two observational catalogues for the 16  
GCs in common.\label{snap-gbn}} 
\end{figure}

The homogeneity of the observational database is a crucial ingredient 
for accurate relative age measurement. In this paper, we use two 
catalogues of CMDs coming from rather different observational systems, 
with similar, but not identical bandpasses. This means that we 
cannot directly compare the parameters obtained from the empirical 
CMDs.  Even if the models used for the age determination have been 
appropriately transformed into the two observational systems before 
comparison with the observed parameters, it is important to verify 
that the ages coming from the two data sets are mutually 
consistent. For this reason, we took advantage of the $16$ clusters in common 
between the HST snapshot and the groundbased data set. 
Figure~\ref{snap-gbn} shows the difference between the ages from 
the snapshot and the groundbased catalogues for the common clusters. 
The average difference is consistent with zero, and there is no trend 
with the cluster metal content. Figure~\ref{snap-gbn} tells us that we 
can merge the two age catalogues, and use them together in  
analysing the relative ages of the Galactic globular clusters. 
 
\subsection{Comparison with other models} 

Comparison with stellar models is critical for any cluster age 
determination. In principle, differences in the input physics
(i.e. equation of state, opacities)  
input parameters (i.e. the helium enrichment ratio $\Delta
{\rm Y}/\Delta {\rm Z}$) and transformations 
from the theoretical to the observational plane, may have a
significant impact on the final ages and the inferred star 
formation history of the Galactic GC system. 
 
\begin{figure}
\includegraphics[width=1.0\textwidth,angle=0]{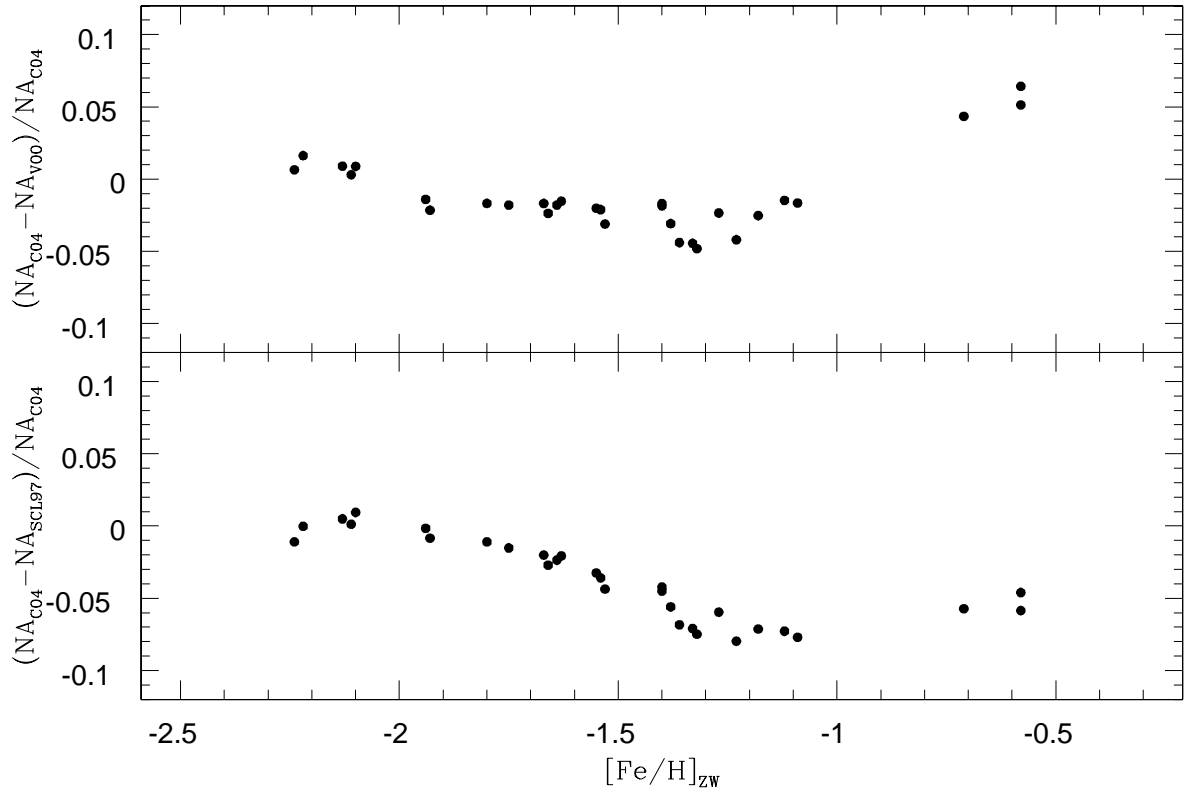}      
\caption{Comparison of the normalised ages (NA) estimated using three different
model databases. C04, SCL97 and V00 denote, respectively, the \citet{cassisi04}, 
\citet{scl97} and \citet{v00} models.\label{confteo}} 
\end{figure}

To assess the consistency of our relative age determination, at 
least within the recent theoretical framework, we compared in 
Fig.~\ref{confteo} our normalised ages for the groundbased sample with the 
values estimated using the two isochrone sets already employed by 
\citet{rspa99}, i.e. the 
Straniero, Chieffi, \& Limongi (1997) and \citet{bv01} isochrones 
calculated by interpolating the evolutionary tracks by \citet{v00}.
Apart from the heavy element mixture, 
these two sets of isochrones differ from our reference models mainly
with respect to the adopted equation of state, bolometric corrections,
and helium abundance at a given metallicity. About this latter
point, the models by \citet{scl97} use ${\rm Y}=0.23$, 
whereas \citet{v00} employ ${\rm Y}= 0.235 +2 {\rm Z}$,
and \citet{cassisi04} make use of ${\rm Y}=0.245+1.4 {\rm Z}$. 
It is worth noticing that the initial He abundance (${\rm Y}$) adopted in the stellar
computation has a strong effect on both the core H-burning lifetime and 
on the HB luminosity level. More in details, for a fixed age, when the value of
${\rm Y}$ increases, the TO of the isochrone becomes fainter, whereas the HB luminosity level
increases.
As a consequence the value of the $\Delta V_{TO-HB}$ parameter increases.
So for a fixed age, the adopted Helium-enrichment ratio, affects the trend of
the $\Delta V_{TO-HB}$ parameter with ${\rm Z}$. The net effect on the age estimate is to make
a cluster of a given metallicity younger when He increases.
Therefore, an increase of the Helium-enrichment ratio enhances, if present, any
age-metallicity 
relationship (in the sense of making the more metal rich clusters younger in
comparison with the more metal poor ones).

Given that \citet{v00} models include a larger Helium enrichment
ratio with respect 
to \citet{cassisi04}, we expect that their use produces younger relative
ages for the more metal rich
clusters. The different Helium-enrichment ratio is also the main reason for the
different behaviours
of the $\Delta V_{TO-HB}$  parameter with the metallicity at a fixed age.

In addition, 
the \citet{scl97} models account for Helium diffusion, which is neglected in the
other isochrone sets.

When deriving the theoretical $\Delta V_{TO-HB}$--age--${\rm [Fe/H]}$ 
calibration from the \citet{scl97} scaled solar isochrones, we used the property that
the $\Delta V_{TO-HB}$ values at a given age are to a good
approximation the same for both a scaled
solar metal distribution and an $\alpha$-enhanced one typical of GCs (see,
e.g. \citealp{scs:93,v00,cassisi04}) with the same
metal mass fraction ${\rm Z}$, at least for
metal poor objects. 
Bearing in mind that the relation between global metallicity ${\rm [M/H]}$ and ${\rm [Fe/H]}$ is different
for $\alpha$-element enhanced mixtures,
the $\Delta V_{TO-HB}$--age--${\rm [M/H]}$ calibration obtained
in this way was then transformed to a $\Delta
V_{TO-HB}$--age--${\rm [Fe/H]}$ relationship using ${\rm [M/H]}={\rm [Fe/H]}+0.3$,
which is appropriate for ${\rm [\alpha/Fe]}=+0.4$. 

The comparison of the normalised ages discloses very good agreement for ${\rm [Fe/H]}_{\rm 
ZW}<-1.9$. Moving towards higher ${\rm [Fe/H]}$ our ages become increasingly
younger up to ${\rm [Fe/H]}_{\rm ZW}\sim-1.1$. The maximum difference with 
the \citet{v00} and \citet{scl97} models is $\sim 5$\% and $\sim 8$\%,
respectively, of the same order of magnitude as the error
bars due to the observational uncertainties. 
For more metal rich clusters the difference decreases to $\sim 5$\%
for the \citet{scl97} models, whereas the difference 
between our ages and those obtained using the \citet{v00} models seems to
be reversed, with the \citet{v00} ages younger than ours by $\sim 5$\% for the
three clusters at ${\rm [Fe/H]}\sim-0.6$. 

We have also compared our normalised ages with the results obtained
using the \citet{g00} models. In this case, there is generally a better 
agreement between the two age sets, with some differences (always smaller 
than $5$\%) at the two extremes of the metallicity interval covered by our 
data set. 
 
Despite these discrepancies, for the purposes of this paper, the 
comparisons in Fig.~\ref{confteo} show that there is a substantial 
agreement among the ages obtained with recent stellar evolution 
models: the overall picture of the age distribution 
discussed in the next Section would not
change if we were to consider the normalised ages obtained from the 
\citet{v00}, \citet{scl97} or \citet{g00} models. It is the 
size of the trends of the average normalised ages with 
metallicity that would marginally change, although an 
agreement at the level of $\sim 5-8$\% (comparable with the 
observational errors) among rather different model sets is 
encouraging. By contrast the trend of the age dispersion with metallicity is 
more solid, as it mainly depends on the dispersion 
of the observed parameters, and much less (and in any case much less 
than the observational errors) on the adopted models. 
 
\section{Discussion} 
 
Figures \ref{figmna2_zw} and \ref{figmna2_cg} plot the normalised ages 
as a function of the metallicity, and of the distance from the 
Galactic centre for the two adopted metallicity scales.  Different 
symbols represent the different metallicity ranges listed in the 
figure caption. Open symbols represent the HST snapshot data, whereas 
filled symbols are for the groundbased clusters. For the clusters in 
common between the two catalogues, we plot the age obtained from the 
vertical parameter measured on the HST snapshot CMDs. The dashed line 
represents the zero point: clusters with a normalised age equal to $1$, 
have an age of $11.2$ Gyr and $10.9$ Gyr for the ZW and CG metallicity 
scales, respectively. The overall trend already displayed by 
Figures~\ref{c04aesnap} and \ref{c04aegb} is more easily visible in 
Figures~\ref{figmna2_zw} and \ref{figmna2_cg}: the most metal poor 
clusters (${\rm [Fe/H]}_{\rm ZW}<-1.7$, ${\rm [Fe/H]}_{\rm CG}<-1.4$) are all 
older, and coeval, with an age dispersion smaller than $0.6$ Gyr, which is 
compatible with a null age dispersion, if we take into account the 
measurement errors. The clusters with $-1.7<{\rm [Fe/H]}_{\rm ZW}<-0.8$ 
($-1.4<{\rm [Fe/H]}_{\rm CG}<-0.8$) are, on average, $1.5$ 
Gyr\footnote{This number would remain the same using the \cite{g00} 
isochrones, but it should be reduced to about $1.0$ and $0.7$ Gyr respectively,
if the models of 
\citet{v00} and \citet{scl97} are adopted.} younger, and show a 
larger age dispersion ($1.0$ Gyr), with an age total range of 
$\sim3$ Gyr.  Interestingly enough,  $15$\% ($5$ out of $32$) of the 
clusters in this metallicity interval seems to be coeval with the most 
metal poor ones.  The age dispersion is independent of the adopted 
stellar evolution model.  More metal rich clusters have on average the 
same age as the intermediate metallicity ones, but apparently with a 
smaller age dispersion, though the small number of objects and the 
differences in the ages from the different model sets 
(cf. Fig~\ref{confteo}) do not allow any firm conclusion for this 
group of clusters.  There is no evidence for a dependence of age on the 
distance from the Galactic centre, though all the clusters with 
$R_{\rm GC}>20$ kpc are old (but these are all metal poor clusters). 
  
In the following we consider a few peculiar objects. On average, we 
note that the metal poor clusters are the oldest 
clusters.  However, among them, two clusters (NGC~4590 and NGC~7078) 
seem to be about $8$\% younger.  Interestingly enough, \citet{yl02} 
listed these two GCs among a group of low-metallicity clusters that 
display a planar alignment in the outer halo, probably the consequence of 
an accretion event from a Galactic satellite system.  On the other 
hand, in the same paper also NGC~7099 and NGC~5024 are indicated as members 
of the same group, while in our study they seem to be coeval with 
the remaining metal poor clusters. 
  
There are a few intermediate metallicity clusters that show a much 
younger age, namely NGC~362 (the youngest of the whole sample, with an 
age $27$\% younger than the zero relative age level), NGC~1261, 
NGC~2808, NGC~3201 and NGC~1851.  All these clusters were already 
known to be younger objects \citep{rspa99}.  Figure 1 in \citet{fp95} displays 
the position of some galactic clusters in galactocentric coordinates 
indicating the position of two planes passing in the vicinity of some 
satellite galaxies of the Milky Way.  NGC~1261, NGC~2808 and NGC~1851 
are located very close to the Fornax-Leo-Sculptor plane. 
The galactocentric coordinates of NGC~362 
and NGC~3201 can be calculated starting 
from the galactic coordinates ($l,\;b$), and the distance from the 
galactic centre \citep[in the revised version of 2003]{harris}: both 
clusters are found to be very close to the same plane. 
  
None of the clusters suspected of being members of the Sagittarius Stream 
\citep{bel03}, and present in our catalogue (NGC~288, NGC~4147, NGC~5053, NGC~5466, 
NGC~5634, NGC~7089), seems to be significantly younger than the average 
of the other clusters with similar metallicity, though it is well 
established that other clusters of the same stream are definitely 
young, e.g. Pal~12 \citep{ros98}, Ruprecht 106 \citep{buo90}, Terzan 7 
\citep{buo95b}. 
 
\section{Conclusions} 
 
We presented a catalogue of homogeneous, relative ages obtained from the 
so called vertical parameter for a sample of Galactic GCs. 
 
The analysis of the relative age distribution gives some hints on the origin of 
the GC system of our Galaxy. 
In particular, we have shown that the low metallicity clusters
(${\rm [Fe/H]}\leq-1.7$) are the oldest GCs.
The low metallicity clusters 
show a very small age dispersion ($<0.6$ Gyr), compatible with a null age dispersion, 
if we account for the measurement errors. 
Among the low-metallicity clusters, there are a couple of objects which 
are marginally younger. These objects are part of a suspected Galactic stream, 
which might indicate accretion from outside the Galaxy, or debris from the 
tidal disruption of a Galactic satellite.  
 
The clusters at intermediate metallicities ($-1.7\le{\rm [Fe/H]}_{\rm 
ZW}\le-0.9$) are up to $3$ Gyr younger than the more metal poor 
ones, with an age dispersion of $\sim1.0$ Gyr. A small sample of 
objects ($15$\%) seems to be coeval with the most metal poor GCs.  It 
must be explicitly noted that, while the age dispersion seems to be a 
solid result, the dependence of the {\it mean} age on metallicity and 
its value are model dependent.  
The empirical facts outlined above imply that most of the GCs with 
${\rm [Fe/H]}_{\rm ZW}\ge-1.7$ were born between $0.5$ and $1$ Gyr after a first 
generation of more metal poor clusters. A few of the intermediate 
metallicity clusters are up to $3$ Gyr younger than the oldest clusters 
in the same metallicity range and this last result is not model 
dependent. Interestingly enough, all these young clusters seem to be 
related to some Galactic streams, i.e. linked to some tidal disruption 
event (of an external object or a smaller satellite), though these 
streams contain also some of the oldest known GCs.  
 
Finally, all the 
metal rich clusters are at least $1$ Gyr younger than the most metal 
poor ones, with a relatively small age dispersion. The metal rich 
sample is still too poor, however, to enable firmer conclusions
to be reached. 
 
We do not see any correlation of the cluster ages with the Galactocentric 
distance. However, all the clusters with Galactocentric distances greater 
than $20$ kpc are old, with a very small age dispersion. These clusters are 
also in the metal poor sample. 

\begin{figure} 
\includegraphics[width=1.0\textwidth,angle=0]{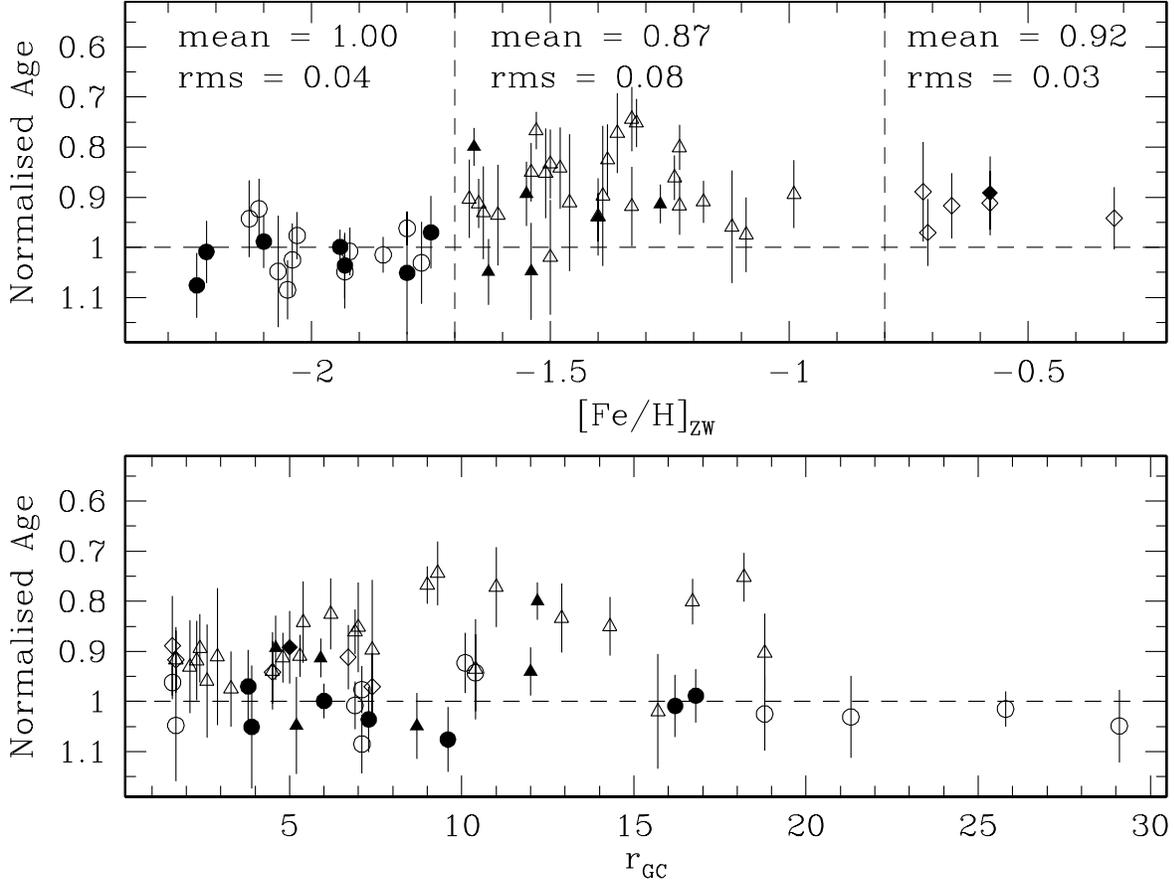}      
\caption{The normalised ages are plotted 
against the ZW metallicity  ({\it upper panel}), and against the distance 
from the Galactic centre  ({\it lower panel}).  Filled symbols are for 
groundbased data and open symbols are for HST snapshot data. 
Circles refer to clusters with [Fe/H]$_{\rm ZW}\leq -1.7$; triangles 
to clusters with $-1.7<$[Fe/H]$_{\rm ZW}<-0.8$, and diamonds to 
clusters with [Fe/H]$_{\rm ZW}\geq -0.8$.\label{figmna2_zw}} 
\end{figure}

\begin{figure}

\includegraphics[width=1.0\textwidth,angle=0]{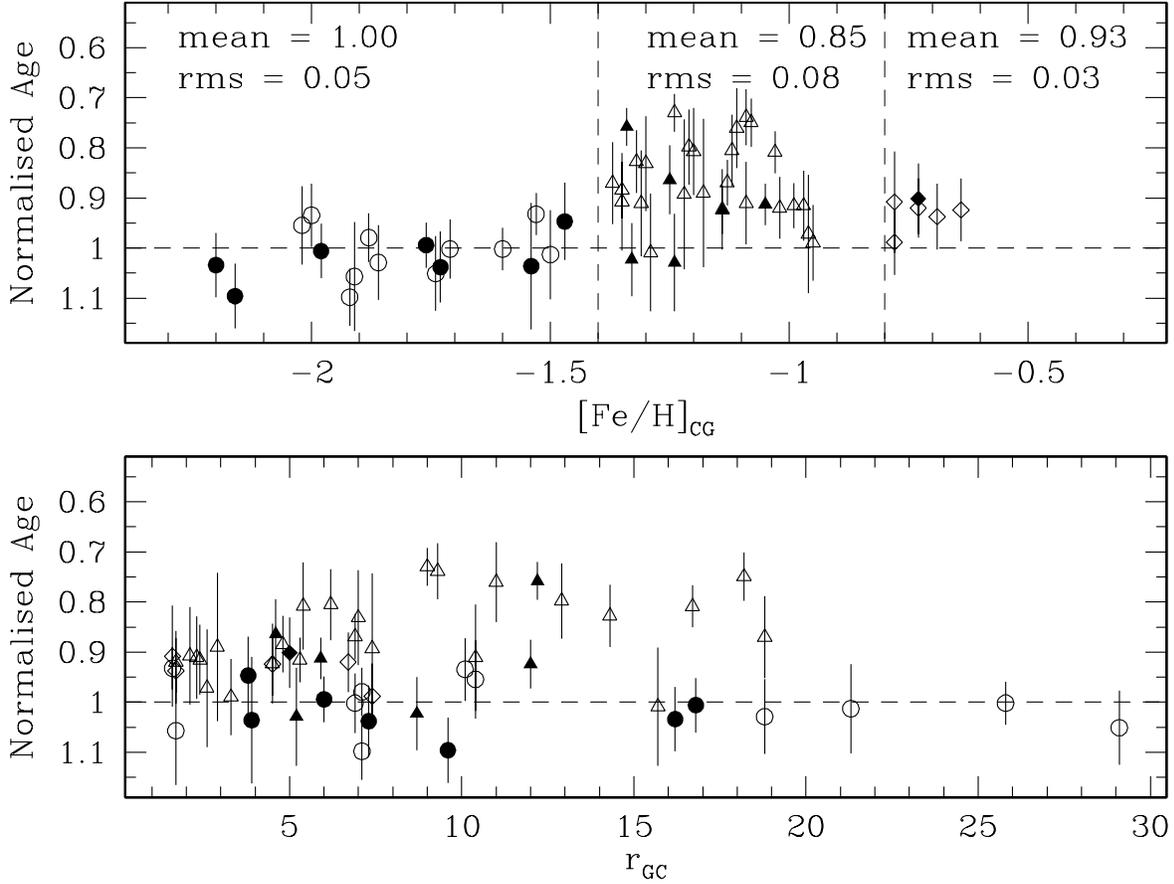}      
\caption{As in Fig.~\ref{figmna2_zw}, but adopting the CG 
  metallicity scale. Circles refer to clusters with [Fe/H]$_{\rm CG}\leq -1.4$; triangles 
to clusters with $-1.4<$[Fe/H]$_{\rm CG}<-0.8$, and diamonds to 
clusters with [Fe/H]$_{\rm CG}\geq -0.8$.\label{figmna2_cg}}     
\end{figure}

In concluding this paper, it is worth adding a final remark. Other
parameters, as Helium variations, ${\rm [\alpha/Fe]}$ differences, 
cluster to
cluster deep mixing (due to internal rotation) affect the vertical
separation between the TO and the ZAHB. We have already commented the
effect of He-enrichment as a function of metallicity in Section
3.3. Here, we note that there is no evidence of significant cluster
to cluster He variations, as shown by Salaris et al. (2004) using our
snapshot database. However, recent results on the massive clusters $\omega$ 
Centauri (Bedin et al. 2004, Piotto et al. 2005) and NGC 2808 (D'Antona 
\& Caloi 2004) indicates that within the same cluster, there could be 
two generations of stars, with the  second generation being Helium enriched. 
If this occurrence is confirmed, and if the same scenario is present also 
in less massive clusters, the ages derived from the vertical parameter 
could be affected.  
As for the other parameters, we plotted our relative ages as a function of
${\rm [\alpha/Fe]}$ for the clusters for which this parameter has been measured,
and we did not find any correlation.  The effect of deep mixing
variations is at the moment impossible to assess from the
observational point of view. In conclusion, the main parameter able to
explain our measured vertical parameter variations, at least inside
the metallicity bins of Figures~\ref{figmna2_zw} and \ref{figmna2_cg} is still age.

\acknowledgments

We warmly thank A.R. Walker and E. Sandquist for providing us their data for some clusters.
We also thank Douglas C. Heggie for a careful reading of the manuscript, 
and the anonymous referee for useful comments.
GP, SC acknowledge partial support from the Ministero dell'Istruzione, Universit\`a  
e Ricerca ({\it MIUR}, PRIN2002, PRIN2003), and from the Agenzia Spaziale Italiana  
({\it ASI}).

\clearpage

\clearpage

\clearpage

\clearpage
 
\clearpage

\clearpage

\clearpage

\clearpage

\clearpage

\clearpage

\begin{table}
\begin{center}  
\caption{Templates for the ZAHB measurement on groundbased clusters.\label{ground}}
\begin{tabular}{ccl}  
\tableline\tableline
ID & Metallicity Interval & CMD Source \\  
\tableline  
NGC 4590 &    $-2.3\le$[Fe/H]$\leq-1.8$           &     \citet{w94}\\ 
NGC 5272 &    $-1.8\le$[Fe/H] $\le-1.5$  &         \citet{r00a}\\ 
NGC 5904 &    $-1.5\le$[Fe/H]$\le-1.3$   &        Sandquist (priv. comm.)\\ 
NGC 1851 &    $-1.3\le$[Fe/H]$\le-1.1$   &        \citet{w98}\\ 
NGC 6362 &    $-1.1\le$[Fe/H]$\le-0.3$            &    Walker (priv. comm.)\\ 
\tableline
\end{tabular}                                                                
\end{center}
\end{table}  

\clearpage

\begin{deluxetable}{cccccccccccccc} 
\tabletypesize{\scriptsize}
\rotate
\tablecaption{Observational parameters and normalised ages for the HST snapshot data. 
Clusters marked with an asterisk show differential reddening in their CMDs.\label{tab1}} 
\tablewidth{0pt}
\tablehead{
ID & [Fe/H]$_{\rm ZW}$ & [Fe/H]$_{\rm CG}$ 
& $r_{\rm GC}$ & \multicolumn{2}{c}{$(F555W)_{\rm TO}\pm 
\sigma$} & \multicolumn{2}{c}{

$(F555W)_{\rm HB}\pm \sigma$} & \multicolumn{2}{c}{ 
$\Delta F555W_{\rm TO-HB}\pm \sigma$} & \multicolumn{2}{c}{Age $\pm$ error} &  
\multicolumn{2}{c}{Age $\pm$ error}
\\
(1) & (2) & (3) & (4) & (5) & (6) & (7) & (8) & (9) & (10) & (11) & (12) & (13) & (14) }
\startdata
IC 4499 & -1.50 & -1.29 & 15.7 &  21.26 &  0.11 &  17.75 &  0.03 &   3.51 &  0.11 &   1.02 &  0.11 &   1.01 &  0.12 \\  
NGC 104 & -0.71 & -0.78 &  7.4 &  17.63 &  0.06 &  14.10 &  0.03 &   3.53 &  0.06 &   0.97 &  0.07 &   0.99 &  0.07 \\  
NGC 362 & -1.33 & -1.09 &  9.3 &  18.75 &  0.07 &  15.50 &  0.03 &   3.25 &  0.08 &   0.74 &  0.06 &   0.74 &  0.06 \\  
NGC 1261 & -1.32 & -1.08 & 18.2 &  20.11 &  0.05 &  16.85 &  0.03 &   3.26 &  0.06 &   0.75 &  0.05 &   0.75 &  0.05 \\  
NGC 1851 & -1.23 & -1.03 & 16.7 &  19.49 &  0.03 &  16.15 &  0.03 &   3.34 &  0.05 &   0.80 &  0.05 &   0.81 &  0.04 \\  
NGC 1904 & -1.67 & -1.37 & 18.8 &  19.61 &  0.08 &  16.25 &  0.03 &   3.36 &  0.09 &   0.90 &  0.08 &   0.87 &  0.08 \\  
NGC 2808 & -1.36 & -1.11 & 11.0 &  19.63 &  0.09 &  16.35 &  0.03 &   3.28 &  0.10 &   0.77 &  0.08 &   0.76 &  0.08 \\  
NGC 3201 & -1.53 & -1.24 &  9.0 &  18.08 &  0.04 &  14.85 &  0.03 &   3.23 &  0.05 &   0.77 &  0.04 &   0.73 &  0.04 \\  
NGC 4147 & -1.77 & -1.50 & 21.3 &  20.46 &  0.09 &  16.98 &  0.03 &   3.48 &  0.09 &   1.03 &  0.08 &   1.01 &  0.09 \\  
NGC 4372 * & -2.03 & -1.88 &  7.1 &  18.94 &  0.05 &  15.58 &  0.03 &   3.36 &  0.06 &   0.98 &  0.05 &   0.98 &  0.05 \\  
NGC 4590 & -2.11 & -2.00 & 10.1 &  19.04 &  0.06 &  15.73 &  0.03 &   3.31 &  0.07 &   0.92 &  0.06 &   0.93 &  0.06 \\  
NGC 4833 & -1.92 & -1.71 &  6.9 &  19.14 &  0.05 &  15.70 &  0.03 &   3.44 &  0.06 &   1.01 &  0.05 &   1.00 &  0.06 \\  
NGC 5024 & -2.04 & -1.86 & 18.8 &  20.27 &  0.08 &  16.83 &  0.03 &   3.44 &  0.08 &   1.02 &  0.07 &   1.03 &  0.07 \\  
NGC 5694 & -1.93 & -1.74 & 29.1 &  22.01 &  0.08 &  18.53 &  0.03 &   3.48 &  0.08 &   1.05 &  0.07 &   1.05 &  0.07 \\  
NGC 5824 & -1.85 & -1.60 & 25.8 &  21.98 &  0.03 &  18.53 &  0.03 &   3.45 &  0.04 &   1.01 &  0.04 &   1.00 &  0.04 \\  
NGC 5904 & -1.38 & -1.12 &  6.2 &  18.49 &  0.07 &  15.15 &  0.03 &   3.34 &  0.08 &   0.83 &  0.07 &   0.81 &  0.07 \\  
NGC 5927 * & -0.32 & -0.64 &  4.5 &  20.25 &  0.05 &  16.79 &  0.03 &   3.46 &  0.06 &   0.94 &  0.06 &   0.92 &  0.06 \\  
NGC 5946 & -1.39 & -1.22 &  7.4 &  21.01 &  0.15 &  17.60 &  0.03 &   3.41 &  0.15 &   0.90 &  0.14 &   0.89 &  0.15 \\  
NGC 5986 & -1.65 & -1.35 &  4.8 &  20.08 &  0.05 &  16.70 &  0.03 &   3.38 &  0.06 &   0.91 &  0.05 &   0.88 &  0.06 \\  
NGC 6171 & -1.09 & -0.95 &  3.3 &  19.23 &  0.07 &  15.70 &  0.03 &   3.53 &  0.07 &   0.98 &  0.07 &   0.99 &  0.08 \\  
NGC 6218 & -1.40 & -1.14 &  4.5 &  18.25 &  0.07 &  14.80 &  0.03 &   3.45 &  0.08 &   0.94 &  0.08 &   0.92 &  0.08 \\  
NGC 6235 & -1.46 & -1.18 &  2.9 &  20.41 &  0.14 &  17.00 &  0.03 &   3.41 &  0.15 &   0.91 &  0.14 &   0.89 &  0.15 \\  
NGC 6266 & -1.23 & -1.02 &  1.7 &  19.76 &  0.05 &  16.30 &  0.03 &   3.46 &  0.06 &   0.92 &  0.06 &   0.92 &  0.06 \\  
NGC 6273 * & -1.80 & -1.53 &  1.6 &  19.95 &  0.04 &  16.55 &  0.03 &   3.40 &  0.05 &   0.96 &  0.03 &   0.93 &  0.04 \\  
NGC 6284 & -1.24 & -1.13 &  6.9 &  20.90 &  0.04 &  17.50 &  0.03 &   3.40 &  0.04 &   0.86 &  0.04 &   0.87 &  0.05 \\  
NGC 6287 & -2.07 & -1.91 &  1.7 &  20.59 &  0.11 &  17.13 &  0.03 &   3.46 &  0.12 &   1.05 &  0.11 &   1.06 &  0.11 \\  
NGC 6342 & -0.66 & -0.69 &  1.7 &  20.55 &  0.05 &  17.07 &  0.03 &   3.48 &  0.06 &   0.92 &  0.07 &   0.94 &  0.07 \\  
NGC 6362 & -1.18 & -0.99 &  5.3 &  18.88 &  0.03 &  15.42 &  0.03 &   3.46 &  0.04 &   0.91 &  0.04 &   0.92 &  0.04 \\  
NGC 6544 * & -1.48 & -1.20 &  5.4 &  18.58 &  0.09 &  15.25 &  0.03 &   3.33 &  0.09 &   0.84 &  0.08 &   0.81 &  0.09 \\  
NGC 6584 & -1.51 & -1.30 &  7.0 &  19.94 &  0.10 &  16.60 &  0.03 &   3.34 &  0.10 &   0.85 &  0.09 &   0.83 &  0.09 \\  
NGC 6637 & -0.72 & -0.78 &  1.6 &  19.50 &  0.09 &  16.04 &  0.03 &   3.46 &  0.09 &   0.89 &  0.10 &   0.91 &  0.10 \\  
NGC 6652 & -0.99 & -0.97 &  2.4 &  19.52 &  0.06 &  16.06 &  0.03 &   3.46 &  0.07 &   0.89 &  0.07 &   0.92 &  0.07 \\  
NGC 6681 & -1.64 & -1.35 &  2.1 &  19.15 &  0.10 &  15.75 &  0.03 &   3.40 &  0.10 &   0.93 &  0.09 &   0.91 &  0.10 \\  
NGC 6717 & -1.33 & -1.09 &  2.3 &  19.24 &  0.07 &  15.80 &  0.03 &   3.44 &  0.08 &   0.92 &  0.08 &   0.91 &  0.08 \\  
NGC 6723 & -1.12 & -0.96 &  2.6 &  19.06 &  0.11 &  15.55 &  0.03 &   3.51 &  0.11 &   0.96 &  0.11 &   0.97 &  0.12 \\  
NGC 6838 & -0.58 & -0.73 &  6.7 &  18.01 &  0.05 &  14.54 &  0.03 &   3.47 &  0.06 &   0.91 &  0.06 &   0.92 &  0.06 \\  
NGC 6934 & -1.54 & -1.32 & 14.3 &  20.28 &  0.06 &  16.95 &  0.03 &   3.33 &  0.07 &   0.85 &  0.06 &   0.83 &  0.06 \\  
NGC 6981 & -1.50 & -1.21 & 12.9 &  20.22 &  0.08 &  16.90 &  0.03 &   3.32 &  0.08 &   0.83 &  0.07 &   0.80 &  0.08 \\  
NGC 7078 & -2.13 & -2.02 & 10.4 &  19.21 &  0.09 &  15.88 &  0.04 &   3.33 &  0.10 &   0.94 &  0.08 &   0.95 &  0.08 \\  
NGC 7089 & -1.61 & -1.31 & 10.4 &  19.44 &  0.10 &  16.03 &  0.03 &   3.41 &  0.11 &   0.94 &  0.10 &   0.91 &  0.11 \\  
NGC 7099 & -2.05 & -1.92 &  7.1 &  18.73 &  0.05 &  15.23 &  0.03 &   3.50 &  0.05 &   1.08 &  0.06 &   1.10 &  0.06 \\ 
\enddata
\end{deluxetable}  

\begin{deluxetable}{cccccccccccccc}  
\tabletypesize{\scriptsize}
\rotate
\tablecaption{Observational parameters and normalised ages for the groundbased data.\label{tab2}}
\tablewidth{0pt}
\tablehead{
ID & [Fe/H]$_{\rm ZW}$ & [Fe/H]$_{\rm CG}$ 
& $r_{\rm GC}$ & \multicolumn{2}{c}{$V_{\rm TO}\pm 
\sigma$} & \multicolumn{2}{c}{ 
$V_{\rm HB}\pm \sigma$} & \multicolumn{2}{c}{ 
$\Delta V_{\rm TO-HB}\pm \sigma$} & \multicolumn{2}{c}{Age $\pm$ error} &  
\multicolumn{2}{c}{Age $\pm$ error}
\\
(1) & (2) & (3) & (4) & (5) & (6) & (7) & (8) & (9) & (10) & (11) & (12) & (13) & (14) } 
\startdata
NGC 104 & -0.71 & -0.78 &  7.4 &  17.65 &  0.08 &  14.10 &  0.03 &   3.55 &  0.08 &   1.00 &  0.10 &   1.02 &  0.10 \\  
NGC 288 & -1.40 & -1.14 & 12.0 &  18.92 &  0.04 &  15.48 &  0.03 &   3.45 &  0.05 &   0.94 &  0.05 &   0.92 &  0.05 \\  
NGC 362 & -1.33 & -1.09 &  9.3 &  18.84 &  0.09 &  15.57 &  0.03 &   3.27 &  0.09 &   0.76 &  0.08 &   0.76 &  0.07 \\  
NGC 1261 & -1.32 & -1.08 & 18.2 &  19.98 &  0.06 &  16.74 &  0.03 &   3.24 &  0.07 &   0.74 &  0.05 &   0.74 &  0.05 \\  
NGC 1851 & -1.23 & -1.03 & 16.7 &  19.58 &  0.07 &  16.23 &  0.03 &   3.35 &  0.07 &   0.81 &  0.07 &   0.82 &  0.07 \\  
NGC 1904 & -1.67 & -1.37 & 18.8 &  19.62 &  0.09 &  16.25 &  0.03 &   3.37 &  0.10 &   0.91 &  0.09 &   0.88 &  0.09 \\  
NGC 2808 & -1.36 & -1.11 & 11.0 &  19.61 &  0.07 &  16.36 &  0.03 &   3.25 &  0.07 &   0.75 &  0.06 &   0.74 &  0.05 \\  
NGC 3201 & -1.53 & -1.24 &  9.0 &  18.11 &  0.05 &  14.83 &  0.03 &   3.28 &  0.06 &   0.80 &  0.05 &   0.77 &  0.05 \\  
NGC 4590 & -2.11 & -2.00 & 10.1 &  19.05 &  0.07 &  15.74 &  0.03 &   3.31 &  0.07 &   0.93 &  0.07 &   0.94 &  0.07 \\  
NGC 5053 & -2.10 & -1.98 & 16.8 &  20.00 &  0.06 &  16.64 &  0.03 &   3.37 &  0.07 &   0.99 &  0.05 &   1.01 &  0.05 \\  
NGC 5272 & -1.66 & -1.34 & 12.2 &  19.00 &  0.04 &  15.76 &  0.03 &   3.24 &  0.05 &   0.80 &  0.04 &   0.76 &  0.04 \\  
NGC 5466 & -2.22 & -2.20 & 16.2 &  19.89 &  0.07 &  16.50 &  0.03 &   3.39 &  0.07 &   1.01 &  0.06 &   1.03 &  0.06 \\  
NGC 5897 & -1.93 & -1.73 &  7.3 &  19.77 &  0.07 &  16.32 &  0.03 &   3.46 &  0.08 &   1.04 &  0.07 &   1.04 &  0.07 \\  
NGC 5904 & -1.38 & -1.12 &  6.2 &  18.47 &  0.03 &  15.12 &  0.03 &   3.34 &  0.04 &   0.83 &  0.04 &   0.81 &  0.04 \\  
NGC 6093 & -1.75 & -1.47 &  3.8 &  19.77 &  0.07 &  16.35 &  0.03 &   3.42 &  0.08 &   0.97 &  0.07 &   0.95 &  0.08 \\  
NGC 6121 & -1.27 & -1.05 &  5.9 &  16.87 &  0.03 &  13.43 &  0.03 &   3.44 &  0.04 &   0.91 &  0.04 &   0.91 &  0.04 \\  
NGC 6171 & -1.09 & -0.95 &  3.3 &  19.17 &  0.06 &  15.68 &  0.03 &   3.48 &  0.07 &   0.93 &  0.07 &   0.95 &  0.07 \\  
NGC 6205 & -1.63 & -1.33 &  8.7 &  18.53 &  0.06 &  15.01 &  0.03 &   3.52 &  0.07 &   1.05 &  0.07 &   1.02 &  0.07 \\  
NGC 6218 & -1.40 & -1.14 &  4.5 &  18.31 &  0.07 &  14.84 &  0.05 &   3.48 &  0.09 &   0.97 &  0.08 &   0.95 &  0.09 \\  
NGC 6254 & -1.55 & -1.25 &  4.6 &  18.48 &  0.05 &  15.11 &  0.05 &   3.37 &  0.07 &   0.89 &  0.06 &   0.86 &  0.07 \\  
NGC 6341 & -2.24 & -2.16 &  9.6 &  18.64 &  0.06 &  15.18 &  0.03 &   3.47 &  0.07 &   1.08 &  0.06 &   1.10 &  0.06 \\  
NGC 6362 & -1.18 & -0.99 &  5.3 &  18.86 &  0.08 &  15.43 &  0.03 &   3.43 &  0.08 &   0.88 &  0.09 &   0.89 &  0.09 \\  
NGC 6366 & -0.58 & -0.73 &  5.0 &  19.14 &  0.06 &  15.70 &  0.03 &   3.44 &  0.07 &   0.89 &  0.07 &   0.90 &  0.07 \\  
NGC 6397 & -1.94 & -1.76 &  6.0 &  16.44 &  0.04 &  13.03 &  0.03 &   3.41 &  0.05 &   1.00 &  0.03 &   0.99 &  0.05 \\  
NGC 6681 & -1.64 & -1.35 &  2.1 &  19.21 &  0.09 &  15.82 &  0.05 &   3.39 &  0.10 &   0.92 &  0.10 &   0.90 &  0.10 \\  
NGC 6723 & -1.12 & -0.96 &  2.6 &  19.07 &  0.09 &  15.53 &  0.03 &   3.54 &  0.09 &   0.99 &  0.10 &   1.01 &  0.10 \\  
NGC 6752 & -1.54 & -1.24 &  5.2 &  17.45 &  0.08 &  13.91 &  0.05 &   3.53 &  0.09 &   1.05 &  0.10 &   1.03 &  0.10 \\  
NGC 6809 & -1.80 & -1.54 &  3.9 &  17.93 &  0.12 &  14.44 &  0.03 &   3.49 &  0.12 &   1.05 &  0.12 &   1.04 &  0.13 \\  
NGC 6838 & -0.58 & -0.73 &  6.7 &  18.01 &  0.06 &  14.52 &  0.03 &   3.49 &  0.07 &   0.95 &  0.08 &   0.95 &  0.08 \\  
NGC 7078 & -2.13 & -2.02 & 10.4 &  19.24 &  0.06 &  15.89 &  0.03 &   3.35 &  0.07 &   0.97 &  0.06 &   0.98 &  0.06 \\  
\enddata
\end{deluxetable}  
  
\end{document}